\documentclass{optica-article}

\journal{opticajournal} % for journals or Optica Open

\articletype{Research Article}

\usepackage{lineno}
\begin{document}

\title{Molecular alignment-assisted spectral broadening and shifting in the near-infrared with a recycled depleted pump from an optical parametric amplifier}
\author{Zhanna Rodnova,\authormark{1} Tobias Saule,\authormark{1} George Gibson,\authormark{1}  and Carlos A. Trallero-Herrero\authormark{1,*}}

\address{\authormark{1}Department of Physics, University of Connecticut, Storrs, CT, 06269, USA\\}

\email{\authormark{*}carlos.trallero@uconn.edu} %% email address is required; see note below about the corresponding author designation

\begin{abstract*} 
We demonstrate how the depleted pump of an optical parametric amplifier can be recycled for impulsive alignment of a molecular gas inside a hollow-core fiber and use such alignment for the broadening and frequency shift of the signal pulse at a center wavelength of $\sim 1300$nm. Our results combine non-adiabatic molecular alignment, self-phase modulation and Raman non-linearities. We demonstrate spectral shifts of up to 204 nm and a spectral broadening of more than one octave. We also report on the time delays at which broadening occurs, which do not coincide with any of the molecular rotational constants. Further, we encounter that maximum frequency shifts occur when the signal and pump have perpendicular polarization instead of parallel.

\end{abstract*}

\section{Introduction}

The invention of lasers in 1959 allowed a variety of applications in science and technology, from spectroscopy \cite{Hansch1971PRL,Smith1971PRL} and communications \cite{Begley2002IEEE} to higher-intensity applications, like studying nonlinear optical processes \cite{Franken1961PRL,Bass1962PRL,Bloembergen1966PRL} and multiphoton ionization \cite{Voronov1965JETPL,Agostini1968IEEE}. After a series of technological breakthroughs in laser design, by the 1980s, the highest achievable peak intensities became limited by self-focusing effects \cite{Luk1989OL}. The invention of chirped pulse amplification (CPA) \cite{Strickland1985OC} by Donna Strickland and Gerard Mourou in 1985 resolved that limitation, paving the way for shorter laser pulses with more bandwidth, allowing for higher peak intensities and enabling a variety of high-intensity applications, from shorter timescale measurements in atomic and molecular physics \cite{Kling2019NatComm,Ott2014Nature,Manzoni2006RSI} and chemical dynamics \cite{Kraus2018NatRevChem,Chang2016IJQC} to attosecond physics \cite{krausz2009RMP,milosevic2004PRL,Ossiander2017NPhys}. 

Applications like these often benefit from frequency conversion from the extreme ultraviolet (XUV) to the near-infrared (NIR). Optical parametric amplifiers (OPAs) are a well-established technology for producing frequency-tunable pulses \cite{Cerullo2003RSI,Ross2002JOSAB,Manzoni2016JO,Brida2010JOA}, and the accessible wavelength range can be further extended with the use of additional non-linear methods.. For example, difference frequency generation (DFG) is a common way of obtaining mid- and long-wavelength infrared femtosecond pulses \cite{Seifert1994OL,Sell2008OL,Andriukaitis2011OL,Lanin2014OL,Lanin2015OL}. In this case, a further wavelength range extension of both the signal and the idler would be useful for concurrently running experiments at a different set of wavelengths, therefore allowing for more flexibility. In another example, degenerate OPAs (DOPAs)can create an extremely broadband gain \cite{Petrov2007OE,Ishii2015JO,Manzoni2016JO}, but are limited in their wavelength range. Extending their wavelength range would allow for experimental applications otherwise not available (i.e., pump-probe experiments or DFG). 

Along with the quest for more frequency tunability, much research has been focused on generating pulses with more bandwidth through spectral broadening. More conventional approaches to spectral broadening use Kerr-induced self-phase modulation (SPM) \cite{Nagy2011OL},\cite{Cardin2015APL} from gas-filled hollow-core fibers (HCFs). In the past, some research has been done on using molecular alignment for spectral modulation \cite{Bartels2002PRL}; recently, this approach has experienced a renaissance of interest. Recent studies investigated optimizing the pulse duration to maximize molecular alignment-induced rotational nonlinearities \cite{Fan2020OL}, \cite{Beetar2020SA}, therefore enhancing
spectral broadening. Others focused on the spectral shaping capabilities of employing molecular alignment for spectral modulation, where a spectral shift was observed around partial rotational wave packet revivals \cite{Mo2022PRA}, with the probe pulse centered around 800 nm achieving a shift of around 100 nm.

An alternative approach to generating larger bandwidth focuses on waveform synthesis, a technique based on spatially and temporally combining multiple pulses with different wavelengths and bandwidths to synthesize pulses with a spectrum spanning multiple octaves \cite{Hassan2016Nature,Wirth2011Science,Hassan2012RSI,Huang2011NatPhot,Mucke2015IEEE,Bahari2020PRA,Liang2017NatComm}. However, implementing such techniques requires a sophisticated feedback mechanism to minimize time jitter between the different arms of the synthesizer. 
OPAs present a natural palette for multiple colors, showing immense potential for waveform synthesis. The depleted pump of an OPA typically possesses poor spatial mode quality \cite{Wilson2014JOSAB} and often gets discarded; however, in the past, our group has shown \cite{Wilson2014JOSAB}, that it can be broadened in combination with the signal. This is advantageous as it produces additional spectral components that are phase-locked with the signal.
In this work, we combine non-adiabatic ($\tau<<T_{rev}$) molecular alignment with SPM- and
Raman-induced spectral broadening. The technique we propose demonstrates a spectral shift of up to 204 nm, as well as broadening enhancement. We show that the recycled depleted pump from an OPA can be effectively used as an alignment pulse, therefore enhancing the overall optical efficiency. In addition, we demonstrate that a depleted pump can be simultaneously broadened with the signal, showing promise for continuous multi-octave spectrum generation; unlike in \cite{Wilson2014JOSAB}, our spectrum contains wavelength components that bridge the spectral gap between the depleted pump and the signal.

\section{Experimental setup}
\begin{figure}[htbp]
\centering
\fbox{\includegraphics[width=0.8\linewidth]{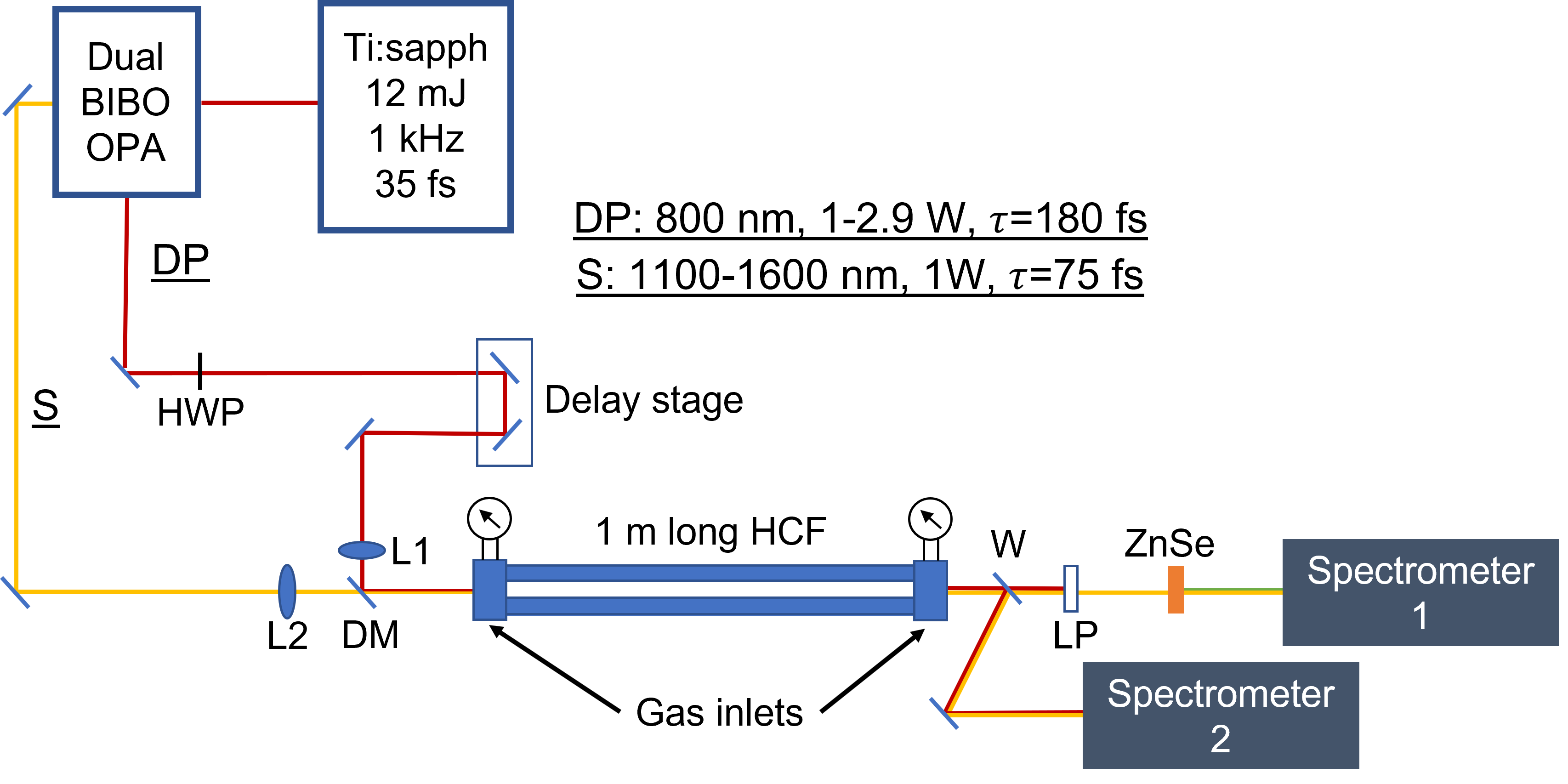}}
\caption{Experimental setup. OPA - optical parametric amplifier; HWP - $\lambda/2$ plate; L1, L2 - lenses for focusing depleted pump (DP) and signal (S) into the fiber, respectively; DM - dichroic mirror, reflecting the DP and transmitting the signal; HCF - 1m-long hollow-core fiber (few-cycle, Inc.); W - wedge; LP - long-pass filter.}
\label{exp_setup}
\end{figure}

Fig. \ref{exp_setup} shows a schematic of the experimental setup.
The technique employs a titanium-sapphire (Ti:sapph) laser system, lasing around 800 nm, with 12 mJ energy per pulse, 1 kHz repetition rate, and pulse duration of 35 femtoseconds (fs). The laser is used to pump an asymmetric dual bismuth triborate (BIBO)-based optical parametric amplifier (OPA). Pump power is split between two arms of the OPA, with 70\% pumping the strong arm and 30\% pumping the weak arm. The OPA generates two signal-idler pairs, with signals tunable from 1080 to 1600 nm, pulse duration of 50-80 fs, and idlers tunable from 1600-2600 nm, pulse duration of 30-100 fs. The power of the strong signal-idler pair is up to 2.7W, corresponding to a 32\% conversion efficiency. Details on the OPA design can be found in \cite{Davis2021OE}. For the experiments presented here, only the strong arm of the OPA was used.
This experiment was performed using the depleted pump (DP) and the strong signal from the OPA tuned to a central wavelength of 1300 nm. In this experiment, the DP, with a pulse duration of 180 fs, was used to align molecules (i.e., N\textsubscript{2}, N\textsubscript{2}O, CO\textsubscript{2}, O\textsubscript{2}) in a hollow-core fiber (HCF). We chose these molecular gases due to their wide availability, their relatively high ionization potential (above 12 eV), and their high nonlinear index of refraction.  The signal has a pulse duration of 75 fs. A $\lambda/2$ plate (half-wave plate, HWP) was used to control the polarization of the signal, with the signal polarization initially perpendicular to the polarization of the DP. Data was taken with polarizations of the DP and the signal parallel as well as perpendicular. A delay stage with a minimum step resolution of 1 $\mu$m was used to delay the pump pulses with respect to the signal pulses. The two beams were spatially recombined, using a dichroic mirror (DM) that reflects the DP and transmits the signal, and coupled into a 1m-long HCF (few-cycle, Inc.) filled with a molecular gas; unless specifically mentioned, a 400 \textmu m-core fiber was used. To ensure coupling to the fundamental EH\textsubscript{11} mode of the HCF, beams were focused to a $\frac{1}{e^2}$ spot size $2\omega=0.7a$, where \textit{a} is the core diameter of the HCF \cite{Cardin2015APL,Abrams1972QE,Marcatili1964Bell}. Gas was supplied into the HCF using gas inlets on both sides of the fiber setup; pressure is measured by two pressure gauges, set on both sides of the setup. All pressures mentioned in the paper are absolute. The fiber setup employed uncoated fused silica entrance and exit windows to ensure both DP and signal were transmitted and to prevent spectral modulation of the output spectrum. As a result of spatial imperfections in the signal mode, coupling losses were the dominating loss mechanism, resulting in a power loss at the front of the HCF. The intensity of the signal inside the HCF was kept around $2 \cdot 10^{13} \frac{W}{cm^2}$ at all times to prevent ionization.
The output of the HCF was split with a wedge, where the transmission of the wedge went through a long-pass filter with a cut-on wavelength $\lambda$=1000 nm to filter out the DP. It then passed through a zinc-selenide (ZnSe) crystal, generating a second harmonic (SH). The SH was necessary due to the wavelength range limitations of spectrometers available during the experiment. Due to its random quasi-matching properties (RQPM) \cite{Baudrier-Raybaut2004Nature,Zhang2019OL,Kawamori2019PRA,Kupfer2018JAP,Davis2022OE}, ZnSe does not require tuning of the phase matching angle of the crystal, making it the perfect candidate for SH generation. The SH spectrum was measured by Spectrometer 1. The reflection from the wedge, containing the signal and the DP, was measured by Spectrometer 2. The delay stage was used to scan across various regions of the rotational structure induced by the DP or signal (depending on the delay), in particular the partial and full revivals (overlap, 
 \textonequarter T\textsubscript{rev}, \textonehalf T\textsubscript{rev}, \textthreequarters T\textsubscript{rev}, T\textsubscript{rev}) to look for changes in the output spectrum. Experimental data were plotted as a two-dimensional map of spectra as a function of the delay between the aligning DP pulse and signal pulse. 

\begin{figure}[htbp]
\centering
\fbox{\includegraphics[width=0.8\linewidth]{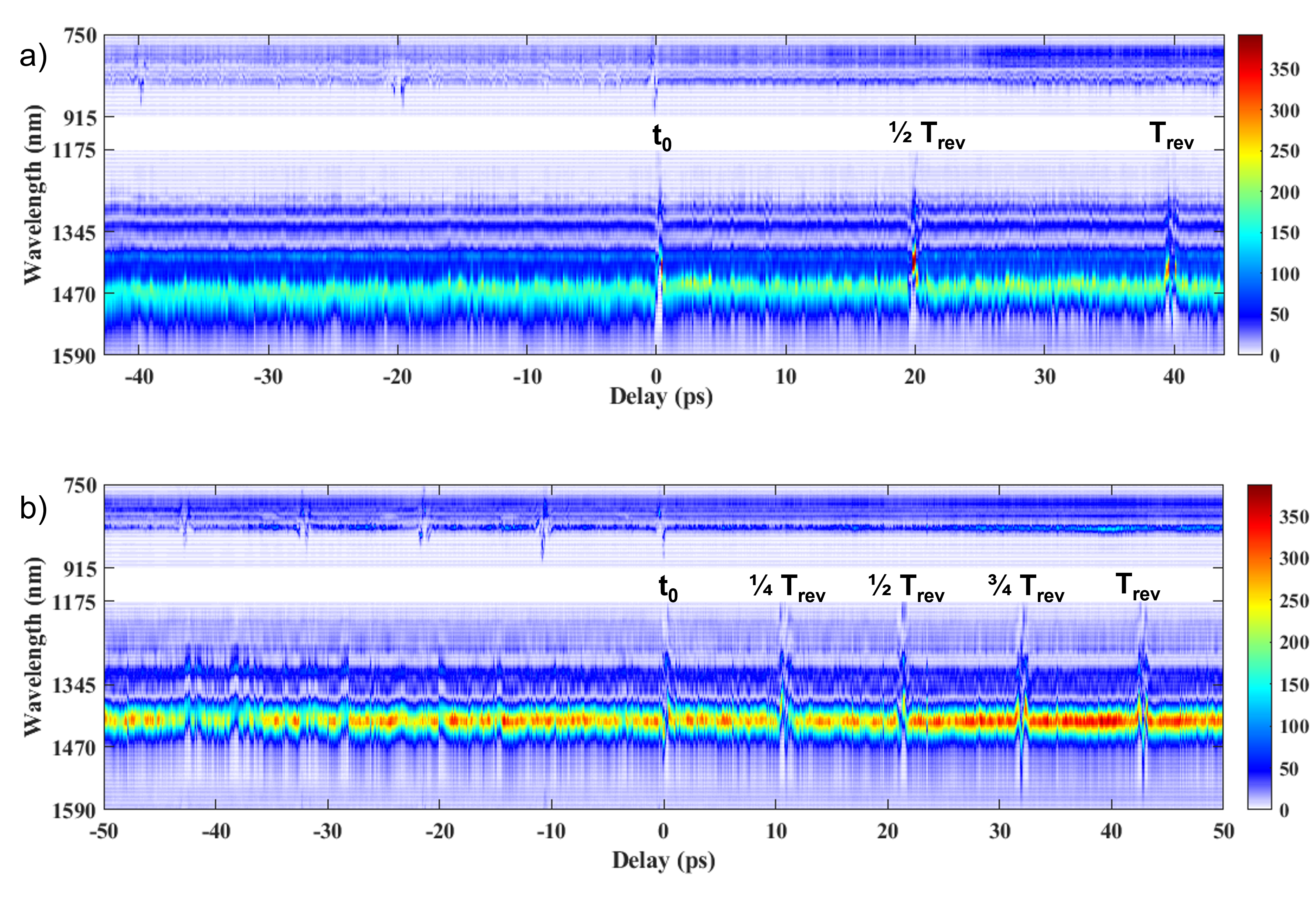}}
\caption{Examples of output spectra obtained by Spectrometer 2. The signal of the OPA was centered around 1300 nm. (a) Output spectrum for 1.36 bar of N\textsubscript{2}O (T\textsubscript{rev}=39.8 ps), signal and DP cross-polarized, (b) Output spectrum for 1.01 bar of CO\textsubscript{2}(T\textsubscript{rev}=42.7 ps), signal and DP cross-polarized.}
\label{traces}
\end{figure}

Fig. \ref{traces} shows two examples of delay-dependent output spectra. Both figures were obtained by performing a scan over -T\textsubscript{rev} to T\textsubscript{rev} for the respective gases. In both figures, at times before overlap (t\textsubscript{0}=0 ps), the signal pulse arrives earlier and has an aligning effect on the DP spectrum. After overlapping, the DP pulse arrives earlier and has an aligning effect on the signal spectrum. To determine the time of overlap accurately, the experiment was performed with the HCF filled with argon (Ar); due to being an atomic gas, argon had a clear overlap region with the DP and signal intensities adding up at t\textsubscript{0} when their polarizations were parallel. Fig.\ref{traces}(a) shows the output spectra from a 1.36 bar nitrous oxide (N\textsubscript{2}O)-filled HCF, whereas Fig.\ref{traces}(b) shows the spectra from an HCF filled with 1.01 bar of carbon dioxide (CO\textsubscript{2}). Data in both figures are taken with signal and DP cross-polarized. In Fig.\ref{traces}(a), a lack of structure at quarter-revivals (\textonequarter T\textsubscript{rev}, \textthreequarters T\textsubscript{rev}) is evident, characteristic for N\textsubscript{2}O, due to the molecule having a 1:1 ratio of even and odd rotational states. On the other hand, Fig.\ref{traces}(b) shows pronounced quarter-revivals as well as the half-revival and full-revival of CO\textsubscript{2}.

\section{Results and discussions}
\begin{figure*}
\centering
\fbox{\includegraphics[width=0.5\linewidth]{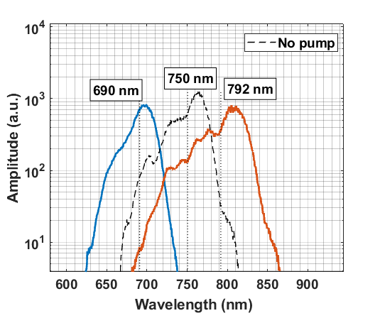}}
\caption{Second harmonic (SH) of the output spectra from a fiber filled with 1.7 bar of N\textsubscript{2}O, signal and DP are cross-polarized. The signal of the OPA was centered around 1300 nm. The dashed line represents the output spectrum without DP. The wavelengths specified are center-of-mass wavelengths for each spectrum. The tunability range of 102 nm in SH corresponds to 204 nm in the fundamental.}
\label{shift}
\end{figure*}

The signal of the OPA was centered around 1300 nm. As the signal interacted with the molecular gas, it experienced spectral broadening due to Kerr-induced SPM, changing its central wavelength. However, when scanning over various delay regions of interest (overlap, 
 \textonequarter T\textsubscript{rev}, \textonehalf T\textsubscript{rev}, \textthreequarters T\textsubscript{rev}, T\textsubscript{rev}), we observed noticeable spectral shifts around the partial revivals along with spectral broadening. The experimental results are summarized in Table \ref{table_shift}, where the largest experimentally observed center-of-mass (COM) wavelength shift is given, together with the polarizability difference for each of the four molecules. The absolute wavelength shift for each molecule was comparable between different partial revivals (¼Trev, ½Trev,
etc.), except in N\textsubscript{2}. In N\textsubscript{2}, the shifts around quarter-revivals (\textonequarter T\textsubscript{rev} and \textthreequarters T\textsubscript{rev}) were half of those around half-revivals (\textonehalf T\textsubscript{rev} and T\textsubscript{rev}).
 \begin{table}
\centering
\begin{tabular}{|c | c | c | c |} 
 \hline
 Molecule & $\Delta \alpha (10^{-25}$ cm\textsuperscript{3}) & COM shift, $\perp$ & COM shift, $\parallel$\\
 \hline
 N\textsubscript{2}O & 28.1$\pm$1.1 \cite{Wahlstrand2012PRA} & 204 nm & 130 nm \\
CO\textsubscript{2}&  21.09 \cite{Le2009PRA} & 136 nm & 110 nm\\ 
O\textsubscript{2} & 10.2$\pm$0.4 \cite{Wahlstrand2012PRA} & 54 nm & 49 nm\\
 N\textsubscript{2} & 6.7$\pm$0.3 \cite{Wahlstrand2012PRA} & 49 nm & 41 nm \\

 \hline
\end{tabular}
\caption{Largest COM wavelength shift and polarizability differences for N\textsubscript{2}O, CO\textsubscript{2}, O\textsubscript{2} and N\textsubscript{2} when signal and DP polarizations were perpendicular ( $\perp$ ) and parallel ($\parallel$). The absolute shift was comparable between different partial revivals for a given molecule, except in N\textsubscript{2}. In N\textsubscript{2}, the shift around quarter-revivals (\textonequarter T\textsubscript{rev} and \textthreequarters T\textsubscript{rev}) was half of that around half-revivals (\textonehalf T\textsubscript{rev} and T\textsubscript{rev}). The number density of molecules is $N=4.2 \cdot 10^{25}$ m\textsuperscript{-3} for all gases, assuming ideal gas law.}
\label{table_shift}
\end{table}
 Figure \ref{shift} shows a wavelength shift that occurred in 1.7 bar of N\textsubscript{2}O, with the COM wavelength of the SH spectrum experiencing a redshift of 42 nm followed by a blue shift of 60 nm as the delay between the DP and the signal increased. This shift corresponded to a tunability range of 102 nm in the SH and 204 nm in the fundamental, which was the largest shift observed. In 1.7 bar of CO\textsubscript{2}, the SH COM wavelength experienced a blue shift of 50 nm followed by a red shift of 18 nm as the delay increased, corresponding to a tunability range of 68 nm in SH and 136 nm in the fundamental. Lastly, the spectral shift in N\textsubscript{2}O was studied with DP and signal polarizations parallel, only showing a COM SH wavelength tunability range of 65 nm, equivalent to 130 nm in the fundamental. The change of the refractive index of a molecular gas is given by \cite{Mo2022PRA}:
 \begin{equation}\label{totaldeltan}
     \Delta n(t)=\Delta n_{K}+\Delta n_{R}+\Delta n_{Align},
 \end{equation}
 where $\Delta n_{K}=n_{2}I(t)$ - instantaneous response due to the Kerr effect, $\Delta n_{R}$ - delayed contribution due to the Raman effect, $\Delta n_{Align}$ - delayed contribution due to molecular alignment.
 When the polarization of the signal is parallel to the polarization of the DP, the change of the refractive index of molecular gas due to molecular alignment is given by \cite{Chen2007OE}:
 \begin{equation}\label{deltanpar}
     \Delta n_{Align}=\Delta n_{\parallel}(t)=\frac{ 2 \pi N}{n_{0}}\Delta\alpha(<cos^2 \theta>_{t}-\frac{1}{3}),
 \end{equation}
 where $\Delta n_{\parallel}(t)$ - change in the nonlinear index in the cross-polarized case, N - number density of molecules, n\textsubscript{0} - linear refractive index, $\Delta \alpha=\alpha_{\parallel}-\alpha_{\perp}$ - the difference between the parallel and perpendicular components of polarizability, $<cos^2 \theta>_{t}$ - expectation value of the thermally averaged alignment. When the polarizations of the signal and the DP are perpendicular\cite{Mo2022PRA,Hoque2011PRA},
 \begin{equation}\label{deltanperp}
     \Delta n_{Align}=\Delta n_{\perp}(t)=-\frac{1}{2}\Delta n_{\parallel}(t).
 \end{equation}Based on Eq. (\ref{deltanperp}), we would expect the wavelength shift to be larger when the DP and signal polarizations are parallel. However, our data consistently showed a larger wavelength shift for the cross-polarized case. We believe that to be because our pulse duration is in a regime where Raman and Kerr contributions to the nonlinear refractive index can no longer be treated and evaluated independently \cite{Wahlstrand2012PRA, Langevin2019PRA}.

%\begin{figure}
%\centering
%\fbox{\includegraphics[width=0.8\linewidth]{broadening_together_v7.png}}
%\caption{(a), (b) Second harmonic (SH) of output spectra from a fiber filled with 1.7 bar of CO\textsubscript{2}, signal and DP have parallel polarization directions. Data were taken around the temporal overlap. (c) SH of output from a fiber filled with 1.7 bar of argon. The dashed line represents the output spectrum without DP.}
%\label{broad}
%\end{figure}
After studying the spectral shift that occurred around partial revivals, we set out to investigate the effect of an aligning pulse on spectral broadening. We anticipated the largest spectral broadening to occur at the exact temporal overlap between the DP and the signal, with DP and signal polarizations parallel and their intensities effectively adding together. However, the biggest broadening did not occur at t\textsubscript{0} but some $\Delta t$ after, indicating a delayed response of the medium that enhanced the spectral broadening. The experimental results are summarized in Table \ref{table_broad}, where the largest experimentally observed bandwidths are given, together with the delays where they were observed. \begin{table}
\centering
\begin{tabular}{|c | c | c | c | c|} 
 \hline
 Molecule  & 10 dB, $\parallel$ & $\Delta t_{\parallel}$ & 10 dB, $\perp$ & $\Delta t_{\perp}$ \\
 \hline
 N\textsubscript{2}O & 417$\pm$3 nm & 406 fs & 330$\pm$15 nm & 180 fs\\
CO\textsubscript{2}& 393$\pm$4 nm & 253 fs & 225$\pm$16 nm & 213 fs \\ 
%O\textsubscript{2} & 243 nm & 100 fs & 236 nm & 80 fs \\
N\textsubscript{2} & 318$\pm$17 nm & 133 fs & 307$\pm$36 nm & 40 fs \\

 \hline
\end{tabular}
\caption{Largest 10 dB bandwidths for spectra from 1.7 bar of N\textsubscript{2}O, CO\textsubscript{2}, O\textsubscript{2} and N\textsubscript{2}, when signal and DP polarizations were parallel ($\parallel$) and perpendicular ( $\perp$ ). The signal of the OPA was centered around 1300 nm. The largest bandwidth was consistently observed in the region of temporal overlap between the DP and the signal. $\Delta t$ is the exact delay in femtoseconds from temporal overlap t\textsubscript{0} where the largest bandwidth occurs.}
\label{table_broad}
\end{table}
Unlike our spectral shift results, we did not see a contradiction with Eq.  (\ref{deltanperp}) in the effect of an aligning pulse on spectral broadening. Our data consistently showed a larger broadening when the DP and the signal polarizations were parallel. The largest broadening enhancement was seen in 1.7 bar of N\textsubscript{2}O, where a 251 nm 10 dB bandwidth was enhanced to 417 nm. In 1.7 bar of CO\textsubscript{2}, the bandwidth without the DP present was 310 nm, enhanced to 393 nm. The experiment was repeated with 1.7 bar of N\textsubscript{2}O, with DP and signal polarizations perpendicular, but that did not yield an increase in bandwidth as significant as in the same polarization case. Lastly, the setup was tested with 1.7 bar of argon to confirm the attractiveness of such a technique for spectral broadening. Ar is one of the more commonly used gases for spectral broadening due to its availability, relatively high ionization threshold (above $10^{14} \frac{W}{cm^2}$ \cite{Augst1991JOSAB}), and high nonlinear refractive index n\textsubscript{2}, resulting in  Kerr-induced spectral broadening due to self-phase modulation. Other commonly used gases, xenon and krypton, are more expensive and generally less available. In our experiment, 1.7 bar of argon showed a 10 dB bandwidth of 265 nm, which was significantly smaller than the enhanced broadening from CO\textsubscript{2} and N\textsubscript{2}O. We were only able to see an enhanced broadening in N\textsubscript{2} when we used a smaller fiber with a 300 \textmu m core diameter. In O\textsubscript{2}, we did not see a strong enough dependence of spectral broadening on the presence of an aligning pulse. We believe using a smaller fiber, and therefore, having a higher peak intensity would have changed that.
\begin{figure}
\centering
\fbox{\includegraphics[width=\linewidth]{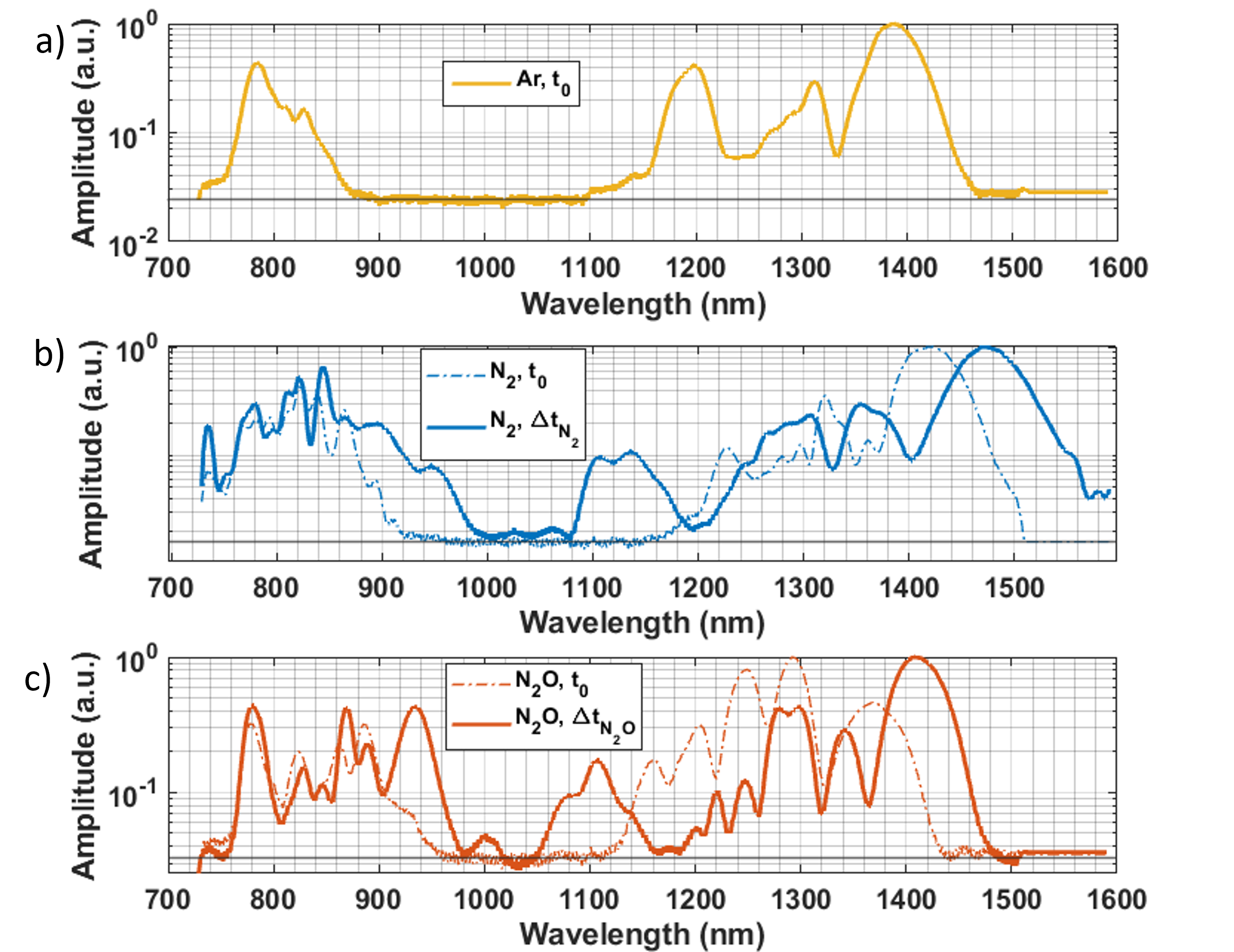}}
\caption{(a) Output spectrum from a 400\textmu m fiber filled with 1.7 bar of Ar. (b) Output spectrum from a 400\textmu m fiber filled with 1.7 bar of N\textsubscript{2}O. (c) Output spectrum from a 300\textmu m fiber filled with 1.7 bar of N\textsubscript{2}. Data in (a) and (b) were taken with the signal centered around 1300 nm. Data in (c) were taken with the signal centered around 1200 nm. The dashed colored line is the spectrum at the time of overlap, t\textsubscript{0}; $\Delta t$ - delay from t\textsubscript{0}. $\Delta t_{N_{2}O}=$406 fs for N\textsubscript{2}O, $\Delta t_{N_{2}}=$133 fs for N\textsubscript{2}. The black line represents background noise.}
\label{cont_spectra}
\end{figure}

After demonstrating the spectral tunability and the spectral broadening capabilities of the setup, we investigated the potential of using the proposed scheme for generating continuous broadband spectra, expecting the enhanced broadening to bridge the spectral gap between the DP and the signal and yield a continuous spectrum. Fig. \ref{cont_spectra} shows output spectra from the HCF filled with Ar (Fig. \ref{cont_spectra}(a)), N\textsubscript{2}O (Fig. \ref{cont_spectra}(b))  and N\textsubscript{2} (Fig. \ref{cont_spectra}(c)). In Fig. \ref{cont_spectra}(b) and (c), the dashed colored spectrum is the spectrum at the time of overlap, t\textsubscript{0}. The spectrum obtained from 1.7 bar of argon (Fig. \ref{cont_spectra}(a)), commonly used for spectral broadening, produces a significant symmetric broadening, characteristic for Kerr-induced broadening. However, using the proposed technique with the HCF filled with 1.7 bar of N\textsubscript{2}O results in a broader spectrum; an even broader spectrum can be obtained using a smaller (300 \textmu m) core diameter fiber filled with 1.7 bar of N\textsubscript{2}. It can be seen that the spectrum slightly after ($\Delta t_{N_{2}O}=$406 fs for N\textsubscript{2}O, $\Delta t_{N_{2}}=$133 fs for N\textsubscript{2})  the overlap is significantly broader, demonstrating the delayed nature of spectral broadening, characteristic of the Raman effect \cite{Fan2020OL}. To obtain a continuous spectrum, we had to either use a smaller fiber, resulting in higher peak intensity (Fig. \ref{cont_spectra}(b)), or lower the signal wavelength to bring it closer to the DP wavelength (Fig. \ref{cont_spectra}(c)). The spectrum in Fig. \ref{cont_spectra}(b) is continuous and spectrum in Fig. \ref{cont_spectra}(c) is nearly continuous. Our resolution is limited by the bit depth of the spectrometer used. We believe our spectrum in Fig. \ref{cont_spectra}(c) is continuous. It can be seen that molecular gases (N\textsubscript{2}O and N\textsubscript{2}) introduce a noticeable red shift of the spectrum, characteristic of Raman-induced nonlinearities \cite{Fan2020OL,Langevin2019PRA}, as well as broadening. In Fig. \ref{cont_spectra}(b) and (c), the output spectrum contains wavelength components around 1100 nm that are otherwise not present. This allows for a continuous spectrum from 729 nm to 1589 nm in N\textsubscript{2} (Fig. \ref{cont_spectra}(b)), spanning more than one octave and showing promise for continuous multi-octave spectrum generation. To confirm our experimental results, we performed theoretical simulations to predict spectral broadening from a gas-filled HCF. The theoretical results agreed with our data in the case of Ar, a noble gas, whereas for molecular gases no such agreement was found.

\section{Concluding remarks}
In conclusion, we have demonstrated a flexible technique that recycles a depleted pump from an OPA to combine non-adiabatic molecular alignment with Kerr-induced self-phase modulation and Raman nonlinearities, allowing for spectral shaping, spectral broadening enhancement, and potential continuous ultra-broadband spectrum generation. We measured  an OPA signal shifting in the center-of-mass (COM) wavelength of the spectrum by up to 204 nm. The COM wavelength shifts observed around different partial revivals were comparable for a given gas, except for N\textsubscript{2}, where the shift around quarter revivals (\textonequarter T\textsubscript{rev} and \textthreequarters T\textsubscript{rev}) was half of the shift observed around half-revivals (\textonehalf T\textsubscript{rev} and T\textsubscript{rev}). We expected the shift to be larger when the DP and the signal polarizations are parallel, but our data consistently showed a larger shift in the cross-polarized case. We believe our pulse duration is in a regime where the changes in the refractive index due to the instantaneous and delayed responses of a molecular medium can no longer be treated separately. We studied the effect of the delayed Raman response of molecular gas on spectral broadening, expecting the contradiction to carry over to broadening as well. However, we found no such contradiction, observing a larger broadening in the case of parallel polarizations. Lastly, we demonstrated the generation of additional spectral components around 1100 nm, bridging the gap between the depleted pump and the signal of an OPA, resulting in a continuous spectrum from 729 nm to 1589 nm, spanning more than an octave without the need for additional jitter control. Adding a broadened spectrum of an OPA idler in the future would allow for extending the spectrum further, making the generation of an ultra-broadband multi-octave spectrum possible.

\begin{backmatter}
\bmsection{Funding}
\bmsection{Acknowledgments} This research was performed under
the Office of Naval Research, Directed Energy Ultra-Short Pulse Laser Division grant N00014-19-1-2339. T. S. was partially funded by AFOSR, Air  Force  Office  of  Scientific  Research grant  FA9550-21-1-0387. Z. R. was partially funded by the Directed Energy Professional Society Graduate Research Grant.

\smallskip

\bmsection{Disclosures} The authors declare no conflicts of interest.

\bmsection{Data Availability Statement} Data underlying the results presented in this paper are not publicly available at this time but may be obtained from the authors upon reasonable request.

\bigskip

\end{backmatter}

% Bibliography
\bibliography{sample}

% Full bibliography added automatically for Optics Letters submissions; the following line will simply be ignored if submitting to other journals.
% Note that this extra page will not count against page length

%%%%%%%%%% If preparing manually:
% \begin{thebibliography}{1}
% \newcommand{\enquote}[1]{``#1''}

% \bibitem{Zhang:14}
% Y.~Zhang, S.~Qiao, L.~Sun, Q.~W. Shi, W.~Huang, L.~Li, and Z.~Yang,
%   \enquote{Photoinduced active terahertz metamaterials with nanostructured
%   vanadium dioxide film deposited by sol-gel method,}
%   {\protect\JournalTitle{Optics Express}} \textbf{22}, 11070--11078 (2014).

% \bibitem{Optica}
% {Optica}, \enquote{{Optica Publishing Group},}
%   \url{http://www.opg.optica.org}.

% \bibitem{FORSTER2007}
% P.~Forster, V.~Ramaswamy, P.~Artaxo, T.~Bernsten, R.~Betts, D.~Fahey,
%   J.~Haywood, J.~Lean, D.~Lowe, G.~Myhre, J.~Nganga, R.~Prinn, G.~Raga,
%   M.~Schulz, and R.~V. Dorland, \enquote{Changes in atmospheric consituents and
%   in radiative forcing,} in \enquote{Climate Change 2007: The Physical Science
%   Basis. Contribution of Working Group 1 to the Fourth assesment report of
%   Intergovernmental Panel on Climate Change,}  S.~Solomon, D.~Qin, M.~Manning,
%   Z.~Chen, M.~Marquis, K.~B. Averyt, M.~Tignor, and H.~L. Miler, eds.
%   (Cambridge University Press, 2007).

% \end{thebibliography}

\end{document}